# X-ray Flares in Orion Low Mass Stars

M. Caramazza[1,2], E. Flaccomio[2], G. Micela[2], F. Reale[1,2], S. J. Wolk[3], and E. D. Feigelson[4]

[1] Dipartimento di Scienze Fisiche ed Astronomiche, Università di Palermo , Via Archirafi 36, 90123, Palermo, Italy  
[2] INAF Osservatorio Astronomico di Palermo, Piazza del Parlamento 1, 90134 Palermo, Italy  
[3] Harvard-Smithsonian Center for Astrophysics, 60 Garden Street, Cambridge, MA 02138  
[4] Department of Astronomy and Astrophysics, Pennsylvania State University, University Park, PA 16802  
email:mcarama@astropa.unipa.it

**ABSTRACT**

*Context.* X-ray flares are common phenomena in pre-main sequence stars. Their analysis gives insights into the physics at work in young stellar coronae. The Orion Nebula Cluster offers a unique opportunity to study large samples of young low mass stars. This work is part of the *Chandra* Orion Ultradeep project (COUP), an ∼ 10 day long X-ray observation of the Orion Nebula Cluster (ONC).
*Aims.* Our main goal is to statistically characterize the flare-like variability of 165 low mass $(0.1 - 0.3 M_\odot)$ ONC members in order to test and constrain the physical scenario in which flares explain all the observed emission.
*Methods.* We adopt a maximum likelihood piece-wise representation of the observed X-ray light curves and detect flares by taking into account both the amplitude and time derivative of the count-rate. We then derive the frequency and energy distribution of the flares.
*Results.* The high energy tail of the energy distribution of flares is well described by a power-law with index ∼ 2.2. We test the hypothesis that light curves are built entirely by overlapping flares with a single power law energy distribution. We constrain the parameters of this simple model for every single light curve. The analysis of synthetic light curves obtained from the model indicates a good agreement with the observed data.
Comparing low mass stars with stars in the mass interval $(0.9 - 1.2 M_\odot)$, we establish that, at ∼ 1 Myr, low mass and solar mass stars of similar X-ray luminosity have very similar flare frequencies.
*Conclusions.* Our observational results are consistent with the following model/scenario: the light curves are entirely built by overlapping flares with a power-law intensity distribution; the intense flares are individually detected, while the weak ones merge and form a pseudo-quiescent level, which we indicate as the characteristic level.

## 1. Introduction

Star formation is often considered a low energy process in which a molecular cloud collapses into a protostar, followed by a phase of quasi-static contraction, eventually leading to the settling of the new born star onto the Main Sequence. However, high energy processes occur as early as the protostellar phase and pre-main sequence (PMS) stars are strong X-ray sources with luminosity up to $10^4$ times the luminosity of the present-day Sun. The source of this high energy radiation is plasma with temperature of $10^6 - 10^8$ K, heated and confined by the stellar magnetic field. X-ray studies of young stars aim to provide insights into the physical origin of the magnetic structures, the plasma heating processes, and the interaction of the X-ray radiation with the circumstellar matter present at the youngest ages. In this context, the study of variability and in particular of stellar flares can provide important insights, in particular into the source of the coronal heating. Although the importance of the magnetic field in transferring energy into the corona is clear, the mechanism by which this energy is transferred into the corona is not established. Numerous heating mechanisms, such as acoustic wave dissipation (Stępień & Ulmschneider, 1989), Alfven wave dissipation (Cheng et al., 1979) and magnetic reconnection phenomena (Parker, 1988; Lu & Hamilton, 1991) have been proposed, all of which might play some role in the overall heating.

Several authors have investigated observationally the last of these proposals, i.e. the possibility that the coronal heating is dominated by small scale flares (see e.g. Audard et al., 1999, 2000, Drake et al., 2000 and references therein), and that the observed X-ray emission is the result of many overlapping flares, only the most intense of which can be resolved in time. According to this hypothesis, originally proposed for the solar case (see Hudson, 1991), the X-ray emission of a star is attributable to flares with a power law energy distribution:

$$\frac{dN}{dE} = k \cdot E^{-\alpha} \qquad \text{with} \qquad \alpha > 0 \qquad (1)$$

where $N$ is the number of flares with energies between $E$ and $E+dE$, emitted in a given time interval. If the index of the power law ($\alpha$) is larger than 2, then a minimum flare energy must be introduced in order to keep the total emitted energy finite and, depending on this cut off, even very high levels of apparently quiescent coronal X-ray emission can be obtained from the integrated effects of many small flares. Direct observational estimates of $\alpha$ are complicated because they require large and unbiased samples of flares. Using various indirect methods, previous authors found $\alpha \sim 2$ for active stars (Audard et al., 1999, 2000; Kashyap et al., 2002; Güdel et al., 2003). Because intense flares are observed to have hard X-ray spectra, a distinguishing characteristic of this microflare scenario would be higher average plasma tem-peratures than the ones seen in the solar corona. This prediction is indeed borne out by observations of PMS stars.

In this context, our aim is to conduct a statistical study of flare properties of the Orion Nebula Cluster (ONC) young stars using data from the *Chandra* Orion Ultradeep Project (COUP). COUP (Getman et al., 2005) is the longest X-ray observation of a stellar cluster ever performed (838 ks). It provides a unique dataset to investigate the X-ray properties of young stars and, thanks to its nearly continuous coverage over 13 days, has enabled statistical studies of variability on large and homogeneous stellar samples, such as those by Flaccomio et al. (2005), Favata et al. (2005), and Wolk et al. (2005). Wolk et al. (2005) in particular studied flare variability on a sample of 27 ~ $1 M_\odot$ ONC members which, by virtue of the dependence of X-ray activity on stellar mass (Preibisch et al., 2005), are also rather strong X-ray emitters. Our principal goal here is to conduct a similar analysis of X-ray variability and flaring on fainter COUP stars with masses $0.1 - 0.3$ $M_\odot$.

In section 2 we define our source sample, describe the method to detect flares and examine the flare properties of our sources; in §4 we introduce a simple model for the X-ray emission and test it through extensive simulations. In §5, we discuss our results.

## 2. Data Analysis

### 2.1. Selection of sources

The X-ray luminosity of young PMS stars depends on stellar mass (Preibisch et al., 2005). For this reason we decided to limit our study to the COUP sources with mass in the interval $0.1 - 0.3$ $M_\odot$, i.e. fully convective M type stars (spectral type: M6.5-M0.5) that will be fully convective M type stars also on the Main Sequence. This selection guarantees a certain homogeneity of our sample in regard to both the physical structure of the stars and their X-ray properties (e.g. count statistics).

Among the 250 COUP sources in the $0.1 - 0.3$ $M_\odot$ mass interval, we will only consider the 165 sources with an effective exposure time longer than 700 Ks and with more than 100 total counts. The first constraint aims at reducing inhomogeneities in source statistics due to the spatial variation of the ACIS-I sensitivity, the second reflects the requirement of a sufficient count statistics for our light curve analysis to be effective. However we have checked that consistent results are obtained if we include weak sources.

Stellar masses were derived by Getman et al. (2005) by comparing the evolutionary tracks of Siess et al. (2000) with stellar positions in the Hertzsprung-Russell diagram, determined from optical spectroscopy and photometry by Hillenbrand (1997). The choice of these particular PMS tracks in Getman et al. (2005) was made for their applicability to a wide range of masses. The use of different theoretical tracks, that have different treatments of convection, degenerate electron pressure and chemical composition would result in a different sample. For example, the tracks by D'Antona & Mazzitelli (1997) give typically lower masses than Siess et al. (2000); their use would increase the number of stars in our sample, including also sources that, according to the tracks by Siess et al. (2000), have masses up to 0.5 $M_\odot$. However, the sources we consider have masses between $0.1 - 0.3$ $M_\odot$ according to both the evolutionary tracks.

### 2.2. Flares Analysis

X-ray flares are characterized by a rapid enhancement of the count-rate followed by a slower decay. Statistical studies of flares require an operative definition of flare that can be used to single them out in the light curves in an unbiased way. In the past, several methods have been used, based on different properties of flares (e.g. Stelzer & Neuhäuser, 2001; Fuhrmeister & Schmitt, 2003). In this work, we use a method similar to that used by Wolk et al. (2005). The procedure consists of dividing the light curve into time intervals during which the count-rate does not vary appreciably, and classifying these intervals (hereafter segments or blocks) according both to their count-rate and to its time derivative.

As in Wolk et al. (2005), we remapped the light curves using the Maximum Likelihood Blocks (MLBs); these are periods during which the count-rate is compatible with being constant at a specified confidence level; their definition is derived from the Bayesian Blocks (Scargle, 1998), but they are based on Maximum Likelihood rather than on Bayesian statistics. The main characteristic of MLBs is that, being computed from the photon arrival times, their temporal length is not based on an *a priori* choice of temporal bin length, but depends on the light curve itself; for this reason MLBs are a useful instrument to quantify different levels of emission, and in particular to detect short impulsive events, that might be missed if we binned the light curves.

The segmentation is performed recursively, first comparing the probability that the photon arrival times derive from an intrinsic constant source with the probability that they derive from two time intervals with different count-rates. If the two segments model is favored above a specified confidence level (we set it to 95%), the same process is repeated recursively on the two segments until no segment can be further divided. In addition, we require that each block includes a minimum of 20 photons.

Our first goal was quantifying variability. We thus grouped the MLBs into broad classes, according to their mean count-rate values. Therefore we needed to establish how much two count-rates have to differ in order to assign them to different classes. We decided that a block count-rate ($R_{block}$) is *compatible* with a count-rate $R$, when $R_{block} \pm \sigma$ is between $R/1.2 - 1.5\sigma$ and $1.2 \cdot R + 1.5\sigma$, where $\sigma$ is the uncertainty associated to the $R_{block}$ value (see below for the choice of the thresholds).

We then singled out three classes of blocks, according to their emission level: *Characteristic*, *Elevated*, and *Very Elevated*.

The *Characteristic Level* ($R_{char}$) was defined as the most frequent emission level in the light curve, i.e. the one that the source keeps for the longest time during our exposure. Indeed, thanks of the exceptional length of the COUP observation, it is often possible to recognize a typical level in the light curves, above which macroscopic flares and other variations of the intensity are visible. We defined $R_{char}$ in the following way. First, we found the count-rate ($R_{long}$) that maximizes the total temporal length of the blocks compatible with it. Then, in order to hold the idea that the characteristic level is a sort of basal emission level, we also included the blocks with:

$$R_{block} < 1.2 \cdot R_{long} + 1.5\sigma \qquad (2)$$

The Characteristic Level ($R_{char}$) is the mean count-rate during the blocks that satisfy the above condition.

The segments classified as *Elevated* satisfy the following condition:

$$1.2 \cdot R_{char} + 1.5\sigma < R_{block} < 1.2 \cdot R_{char} + 2.5\sigma \qquad (3)$$

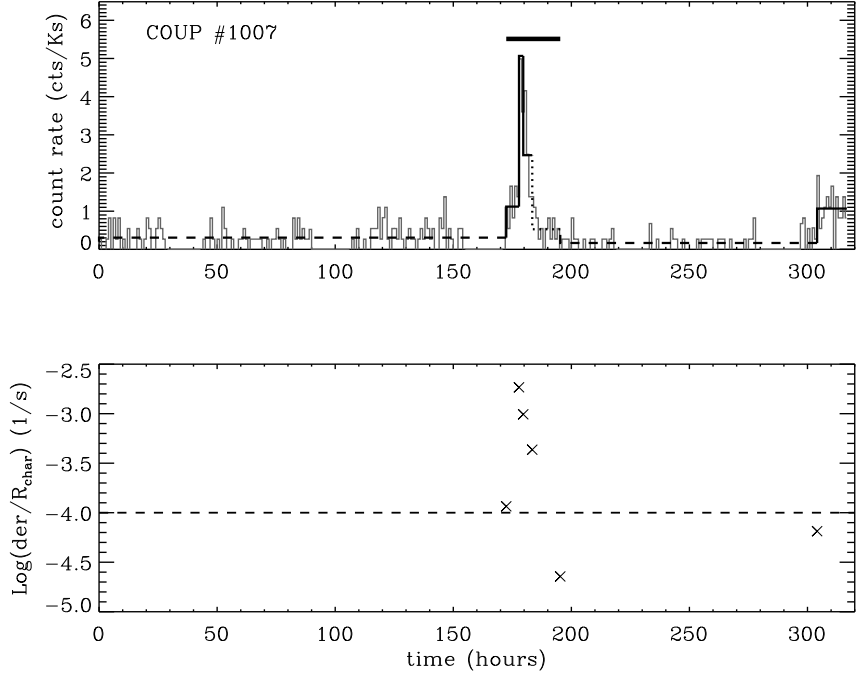

**Fig. 1.** Light curve of COUP source 1007, elucidating our definition of flare. Top panel: On the light curve, represented by an histogram with 1 hour bins, we plot the MLBs. The characteristic blocks are plotted with a dashed line, elevated blocks with a dotted line and the very elevated ones with a solid line. Flaring blocks are identified with a thick horizontal line toward the top of the panel. Bottom panel: $(dR/dt)/R_{char}$, calculated at the interface between blocks. The dashed line represents our threshold, $(dR/dt)/R_{char} = 10^{-4}$. Note that rapid variations of flux are needed to identify a flare: the last block is very elevated but the derivative is under the threshold, therefore the event is not detected as a flare.

These blocks are only slightly above the characteristic level and are not usually associated with macroscopic flares.
Finally, the *Very Elevated* blocks are defined by:

$$R_{block} > 1.2 \cdot R_{char} + 2.5\sigma \qquad (4)$$

In order to define flares, we measured the rate of variation of the photon flux, through the time derivative: $dR/dt$, where $dR$ is the difference between the count-rate of two successive blocks ($R_1$, $R_2$), and $dt$ the minimum value of the temporal length of the two blocks ($\Delta t_1$, $\Delta t_2$).

$$\frac{dR}{dt} \equiv \frac{|R_1 - R_2|}{min(\Delta t_1, \Delta t_2)} \qquad (5)$$

Following Wolk et al. (2005), we noted that flares identified by eye are characterized by:

$$\frac{dR}{dt} \cdot \frac{1}{R_{char}} > 10^{-4} s^{-1} \qquad (6)$$

Using the previous tools, we give the following operative definition: *a flare is a sequence of non-characteristic blocks (elevated and very elevated) beginning with a very elevated block and whose derivative (5) is above our threshold ($R_{char} \cdot 10^{-4} s^{-1}$).*

In figure 1, we illustrate our method: on the basic light curve, with uniform one hour time bins, we overplot the MLBs. Characteristic blocks are plotted with a dashed line, elevated blocks with a dotted line and very elevated ones with a solid line. In the bottom panel, we plot $(dR/dt)/R_{char}$ at the interface between adjacent blocks. The sequence of blocks belonging to the detected flare is indicated by a horizontal line in the upper part of the light curve panel.
The thresholds used in the above definitions were chosen in order to recognize significant flares. Indeed, only macroscopic flares, characterized by a luminosity rise by more than of 2.5 $\sigma$ above 120% of the characteristic level, can be detected. Weaker events are not detected and will be either classified as elevated blocks or blended in the characteristic level. Moreover, the threshold of the derivative is fixed with the purpose of identify events with a steep count-rate raise and it is calibrated on the evident flares recognized by eye in the light curves.

The main difference between our flares detection method and the one used by Wolk et al. (2005) is in the segmentation process: while our algorithm tests at every step the two segments hypothesis, in Wolk et al. (2005) the algorithm, for segments with less than 2000 counts, also tested the probability that the arrival times derive from a three segments model; this extra step improves the sensitivity to faint impulsive events at the expense of an increased computation time. Although the three segments method is more efficient in finding weak flares, the processing time is considerably longer than for the two segments one: since in this work we analyze a large number of simulated light curves, we decided to use a version of the MLB algorithm that only tests the two segments hypothesis, even for segments with less than 2000 photons. However the final results do not depend on this choice.
A further difference is in the choice of the thresholds that here are lower, in order to single out weaker flares in our sources. Moreover, according to our definition, flares start with the first high derivative block, while Wolk et al. (2005) also included all the previous elevated and very elevated blocks. We noted that for our sample this second choice tends to overestimate the duration of flares, and thus decided to modify the definition as described, explicitly incorporating the idea that flares start with a rapid rise.

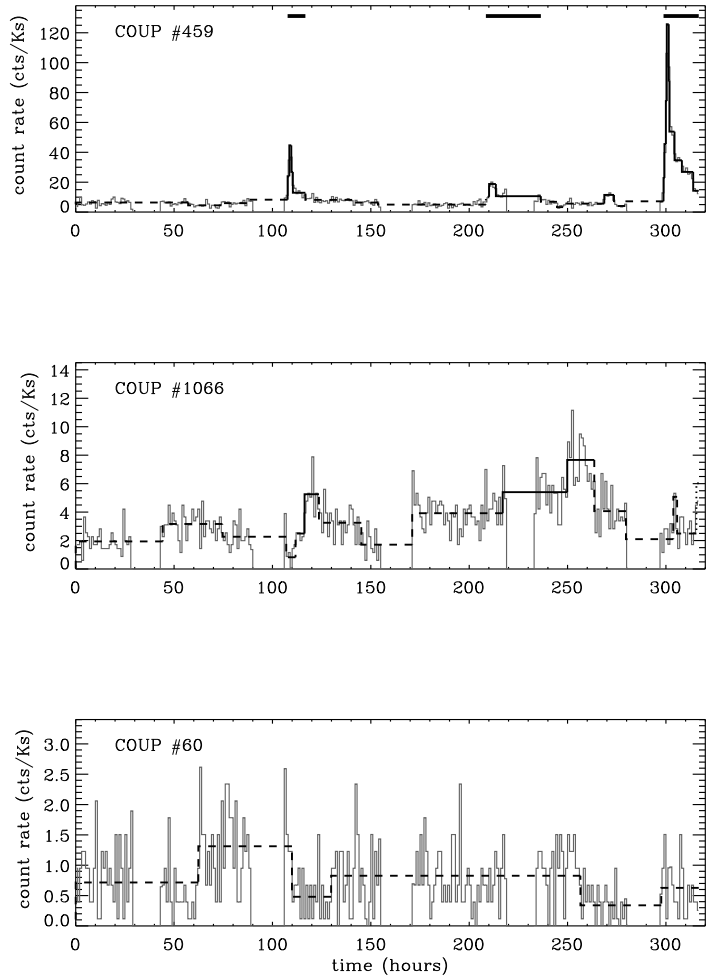

**Fig. 2.** Examples of different types of light curves. Lines and symbols as in Fig. 1. *Top*: light curve of a source with several flares (COUP source 459). *Center*: light curve of a source without detected flares (COUP source 1066). Although there are very elevated blocks, their derivative is under our threshold. *Bottom*: "Constant" light curve (COUP source 60): all the blocks are compatible with the characteristic level (0.8 $cts$ ks$^{-1}$).

We apply the method outlined above to analyze the light curves of the COUP low mass stars. In our sample several kinds of variability are visible. There are sources showing several flares, e.g. source 459 (see first panel of Fig. 2); sources showing considerable variations (there are very elevated blocks), but with no detected flares as $(dR/dt)/R_{char}$ is always below the threshold, e.g. source 1066 (second panel of Fig. 2). Finally there are six sources, like source 60 (third panel of Fig. 2) that show low amplitude variability, all their blocks being compatible with $R_{char}$.

## 3. Observational Results

### 3.1. Flare Frequency

We detected a total of 151 flares in our sample of 165 sources, i.e. approximately one flare per source. The distribution of the number of flares per source, shown in Fig. 3, is not uniform: about 40% of the stars have no flares and there are stars with up to four flares during the observation.

Since the sensitivity of our flare detection method depends on source statistics, we calculated the mean number of flares per source ($N_{flares}$) considering several subsamples of sources, comprising sources with increasing minimum number of counts. The results are reported in Table 1 and plotted in Fig. 4: as expected the mean number of flares increases with increasing source intensity. Uncertainties on the $N_{flares}$ values were calculated assuming Poisson statistics. The observed trend of $N_{flares}$ is most likely the result of a bias introduced by photon statistics, rather than an intrinsic property of the stars.

We compared our results for low mass stars with similar ones obtained for more massive stars, taking into account source statistics. We selected the 27 COUP stars with masses in the range $0.9 - 1.2 M_{\odot}$ studied by Wolk et al. (2005). Since our flare definition is slightly different, we have repeated the analysis of these sources, identifying 46 flares, to be compared to the 41 found by Wolk et al. (2005). The average number of flares per source in solar mass stars ($N_{flares} = 1.70 \pm 0.03$) is higher than for our sample. However, if we restrict our comparison to stars with sim-ilar statistics, as indicated by the filled symbols in Fig. 4 [1], we note that the two samples actually have indistinguishable flare frequencies.

### 3.2. Intensity Distribution

We measured the intensity of a flare as the number of photons emitted during the flare:

$$C = \sum (R_{block} - R_{char}) \cdot T_{block} \quad (7)$$

where $T_{block}$ is the temporal length of each block and the sum is extended to all the blocks of a flare. Note that since we are analyzing a sample of stars at the same distance, this number of photons is related to the released energy, but is not strictly proportional to it, because we have neglected flare-to-flare differences in the X-ray spectra. Figure 5 shows the cumulative distribution function (CDF) of the intensity of flares in our sample. For high counts the distribution is well described by a power law, but it progressively flattens towards low counts, probably because not all low-count flares are detectable.

Following the microflare hypothesis (see eq. 1), we described the high count part of the differential distribution of flare counts as a power law, with index $\alpha$:

$$\frac{dN}{dC} = k \cdot C^{-\alpha} \quad \text{with} \quad \alpha > 0 \quad (8)$$

The cumulative distribution is then also a power law:

$$CDF = \int_C^\infty \frac{dN}{dC} dC = \frac{k}{\alpha-1} \cdot C^{-(\alpha-1)} \quad (9)$$

We have determined the cutoff number of counts $C_{cut}$ above which the observed distribution is compatible with a power law and the relative index $\alpha-1$, with the same method used by Stelzer et al. (2006) (see also Crawford et al., 1970) :

$$\alpha - 1 = 1.2 \pm 0.2 \qquad C_{cut} = 500$$

In agreement with our assumption on the power law shape of the distribution, taking $C_{cut}$ larger than the chosen cutoff, the best fit value of $\alpha$ remains stable within uncertainties, while under this threshold we cannot neglect the incompleteness effect. Since more intense flares are typically hotter, it is not obvious that the slope of the flare count distribution is the same as that of the flare energies. We tested the effect of the different X-ray spectrum of flares on the energy distribution, assuming that the temperature of the flaring plasma varies in the range 3-15 keV, a typical temperature range for flares, proportionally with the the logarithm of the number of photons [2]. The resulting slope of the energy distribution is compatible with the one we have calculated for the number of flares photons. With this conversion a flare with 500 photons has a typical energy of $\sim 10^{34}$ erg, comparable with that of a large solar flare.

## 4. Simulations

In this section we will test the hypothesis that the observed light curves can be reproduced by a model consisting solely of flares with a single power law distribution of flares energies. We will describe the model and relate the free parameters to the observable characteristics of the COUP light curves. Using this model, we will simulate 100 light curves for each star of our sample. The analysis of the synthetic light curves will allow us to test the ability of our analysis method to successfully recover the underlying power law index and to confirm that the observed shape of the CDF is consistent with the hypothesis of a single power law, i.e. that the turnover observed in the distribution in Fig. 5 can be attributed to undetected low-count flares.

### 4.1. Model

According to the microflare hypothesis (see eq. 1) we suggest the following simple model: light curves are formed entirely by overlapping flares, with intensities sampled from a power law distribution with the same index for all the sources:

$$\frac{dN}{dC} = k \cdot C^{-\alpha} \qquad \alpha > 0 \quad (10)$$

with C values in the interval $[C_{min}, C_{max}]$. Assuming that $\alpha > 2$ we can take $C_{max} = \infty$ while keeping the total number of flares and the total number of photons finite; in practice, in our simulations, for numerical reasons, we set $C_{max} = 10^{30}$. From eq. 10, we can calculate the average number of flares occurring during a given time interval, and the corresponding average number of emitted photons for each source:

$$<N_{fl}> = \int_{C_{min}}^{C_{max}} \frac{dN}{dC} dC \simeq \frac{k \cdot C_{min}^{-(\alpha-1)}}{\alpha-1} \quad (11)$$

$$<N_{ph}> = \int_{C_{min}}^{C_{max}} \frac{C \cdot dN}{dC} dC \simeq \frac{k \cdot C_{min}^{-(\alpha-2)}}{\alpha-2} \quad (12)$$

For the sake of simplicity, we model flares as having an instantaneous rise and an exponential decay with an e-folding time $\tau$. The model is then completely specified by four parameters: $\alpha$, $\tau$, $C_{min}$ and $<N_{fl}>$. We decided to fix $\alpha$ and $\tau$ to typical values determined from the observations; $C_{min}$ and $<N_{fl}>$ were instead treated as free parameters and determined from the data for each source. In particular, we chose to relate $C_{min}$ and $<N_{fl}>$ to two observed characteristics of the light curves: the total number of detected photons ($N_{ph}$) and the characteristic level ($R_{char}$). Combining eq. 11 and 12, we obtain:

$$<N_{ph}> = <N_{fl}> C_{min} \frac{\alpha-1}{\alpha-2} \quad (13)$$

This relationship implies that the same $<N_{ph}>$ can be realized with many weak flares (high $<N_{fl}>$ and low $C_{min}$) or with few more intense flares (low $<N_{fl}>$ and high $C_{min}$). We note that $<N_{ph}>$ is independent of $\tau$.

Since the definition of $R_{char}$ is not analytic, we resorted to simulations in order to relate it to model parameters. We first performed an exploratory set of simulations, using the grid of parameter values shown in Table 2. The goal of these simulations is the determination of the relationship between the observables ($R_{char}$ and $N_{ph}$) and the input model parameters ($<N_{fl}>$ and $C_{min}$). We used three different values of $\alpha$ in order to explore the interval suggested by the data analysis, and $C_{min}$ and $<N_{fl}>$ values to cover the observed range of $N_{ph}$ for our sources (see eq. 13 ). Regarding $\tau$, observed flares have decay times in a wide range of values. For simplicity we here chose two values that appear to bracket most flare decay times.

---

[1] In the solar mass case, because of the limited number of sources, we can study only four subsamples.

[2] The count-to-energy conversion factor was computed with the *Portable Multi Mission Simulator* (PIMMS), assuming the interstellar absorption values given by Getman et al. (2005)

Our code builds the synthetic light curves in the following way: it draws the number of counts for each flare from the distribution in eq. 10, randomizes the obtained values according to a Poisson statistical distribution and then distributes the flare start times randomly during the observation time; finally, it generates the photon arrival times for each flare with a decaying exponential distribution with e-folding time $\tau$. Temporal gaps in the observation are reproduced by removing from the final photon list those that fall in these gaps.

We analyzed the simulated light curve with the same method we used for our sources, calculating $R_{char}$, and identifying macroscopic flares, as in Sect. 2.2. In figure 6, we plot the mean $R_{char}$ versus $<N_{fl}>$ for different value of the other parameters, as resulting from 1000 simulations of each model. We note that $R_{char}$ increases with increasing $<N_{fl}>$, $C_{min}$ and $\alpha$. The pairs of curves with the same line style refer to the two different values of $\tau$; since the dependence of $R_{char}$ on $\tau$ is relatively weak, we decided to neglect it, for the moment.

In the light of this analysis, we note that, after fixing $\alpha$ and $\tau$, $C_{min}$ and $<N_{fl}>$ determine the distribution of $R_{char}$ and $N_{ph}$ (with $<N_{ph}>$ given by eq. 13 ). In the following, the power law index is fixed to the measured value $\alpha = 2.2$ (see Fig. 5). Since $\tau$ has no significant influence on $N_{ph}$ and $R_{char}$, we fixed it to the reasonable value of 10 hours: we will further discuss this choice in Sect. 4.3.

### 4.2. Determination of model parameters for the low mass COUP sources

In order to test our simple model, we used the observed $N_{ph}$ and $R_{char}$ to determine Maximum Likelihood estimates of $C_{min}$ and $<N_{fl}>$ for every source in our sample. We then simulated the light curves with our model and compared the characteristics of the simulated and observed data.

In order to establish the relation between the observed $N_{ph}$ and $R_{char}$ and the model parameters $C_{min}$ and $<N_{fl}>$, we generated synthetic light curves, varying $C_{min}$ in the range 0.001-90 and $<N_{ph}>$ in the range 100-10000, with a 0.5 logarithmic step [3]. We simulated 2000 light curves for each of the 55 points in the $C_{min}$, $<N_{fl}>$ grid and analyzed these simulated light curves in the same way of the real light curves, calculating in particular $R_{char}$ and the mean count-rate $R_{mean} = N_{ph}/T_{exp}$, where $T_{exp}$ = 838 ks is the exposure time of the COUP observation.

In figure 7 we plot $R_{mean}$ versus $R_{char}$ for the 5 cases with different $<N_{ph}>$ and $C_{min} = 0.01$ (each point refers to a single simulated light curve). The five samples are represented by relatively concentrated clouds of points, and are easily distinguishable because the mean count-rate and $R_{char}$ increase with increasing $<N_{ph}>$. We note also that, for each $<N_{ph}>$ the distributions of $R_{char}$ are roughly symmetrical but the distributions of the mean count-rate are remarkably skewed toward high values. For this reason we chose to represent each set of simulations with the mean value of $R_{char}$ and the median value of $R_{mean}$ (instead of the mean).

Figure 8 shows the relation between mean $R_{mean}$ and median $R_{char}$ for all the simulation sets. The connected crosses refer to the same value of $C_{min}$ (the error bars refer to the 68% quantiles of the distributions), while the dots represent the real sources in our sample. Since a difference between $R_{mean}$ and $R_{char}$ means that the source is variable, in this plot the more a point is distant from the bisector, the larger the variability of the represented light curve.

We note that the degree of variability of a source is determined more by $C_{min}$ than by $<N_{fl}>$ : sources with a small value of $C_{min}$ are near the bisector, with $R_{mean}$ similar to $R_{char}$, therefore they only experience a low level of variability. Instead, for a given $C_{min}$, $<N_{fl}>$ fixes the intensity of the source.

We estimated $C_{min}$ and $<N_{fl}>$ for each of our sources, interpolating from the grid showed in Fig. 8. In this way, we were able to simulate each source and to test the agreement between the synthetic and the real light curves. Note that if we had represented the simulations with the mean $R_{mean}$ instead of the median, we would have obtained systematically larger values of $C_{min}$ and $<N_{fl}>$, with the result that model light curves would have been also systematically biased toward higher counts than observed.

### 4.3. Simulations of the COUP low mass sample

Using the model parameters obtained in the previous section, we simulated the light curve of every single low mass source in our sample 100 times, and analyzed the light curves with the same method as for the real sources. The simulations qualitatively reproduce the main characteristics of the real light curves: in Fig. 9 we plot, as an example, one of the simulations of sources 459, to be compared with the observed light curve in Fig. 2.

Figure 10 shows the CDFs obtained from each of the 100 simulations of the sample (grey lines) and the one obtained from the observed lightcurves (black dots). We note that the simulated CDFs have a shape similar to that of the observed one, with a power law at high counts and a saturation at low counts. The simulated CDFs have a large scatter at high counts, where they are determined only by few flares. The observed CDF for our low mass stars appears compatible with the simulations.

For each of the 100 simulation sets we repeated the analysis of the flare intensity distribution described in Sect. 3.2, estimating the $\alpha - 1$ and $C_{cut}$ values that best describe the power-law tails. Figure 11 shows the histogram of the $\alpha - 1$ values obtained from the simulations. The distribution has a mean value $\alpha - 1 = 1.2 \pm 0.2$, compatible with the input value. Note moreover that the reported 1 $\sigma$ dispersion in the $\alpha - 1$ values from the simulations is the same as the statistical uncertainty derived from the real data. This supports the validity of our model and our hypothesis that although we are not able to detect all the flares, we do detect enough events in order to estimate the power law slope.

As a further test of the model, we compared the flare frequencies from the simulations with those from the observed data. Figure 12 shows the mean number of simulated and observed flares per source as a function of the number of source counts. For the simulated data, horizontal bars indicate the considered count intervals and vertical bars the 1 $\sigma$ dispersion in the simulation results. We note that the number of flares predicted by the model is in all cases compatible with that of the flares detected on our sources. Nevertheless, we also note that the number of flares of the simulated light curves seem to be systematically higher for the sources in the three highest count bins. In order to explain why our simulated sources appear more variable at high counts, we explore the effect of the only parameter we fixed arbitrarily, i.e., $\tau$.

We noted in Sect. 4.1 that the parameter $\tau$ has a little influence on the determination of $R_{char}$ and $<N_{ph}>$. However, since $\tau$ is the decay time of the flares that form the light curve, it is very unlikely that it has no weight in the model. Because our ability to detect and resolve individual flares depends, for a given flare intensity, on the duration of the flare, $\tau$ will influence the frequency of detected flares, in particular the frequency increases with decreasing $\tau$. We have not enough counts to determine $\tau$ for all flares; we thus tried to verify that our choice is compatible with the observed flares by investigating the relation between $\tau$ and the duration of the flares detected in the simulations, defined as the total temporal length of the blocks associated with the flare.

---

[3] Note that for a given $C_{min}$, in the light of the relation 13, varying $<N_{ph}>$ is equivalent to varying $<N_{fl}>$.

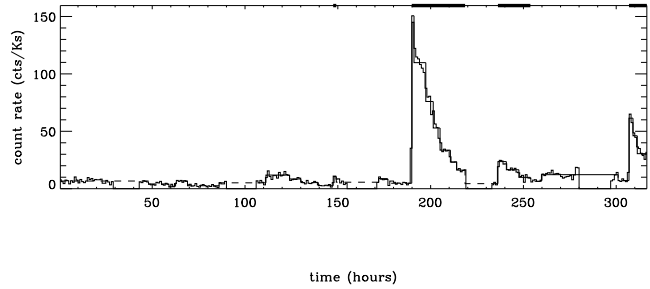

**Fig. 9.** One of the simulations of COUP source 459. The simulations reproduce, on average, the $R_{char}$ and $R_{mean}$ of the real source. See the first panel of figure 2 for a comparison with the observed light curve.

We thus repeated the simulations using the same values of $C_{min}$ and $<N_{fl}>$ as above, for different values of $\tau$ and calculated for the simulated light curves the duration of the detected flares. Table 3 lists the mean duration of detected flares for 5 different values of $\tau$ from 5 to 15 hours. For the real sources the mean temporal length is 21 hours corresponding to $\tau = 10$, supporting our initial choice.

Figure 13 shows the histograms of the duration of flares for the data and three simulation sets (5, 10 and 15 hours). We note that the distribution of the duration of flares for the data is not well described by a single $\tau$ representation: short flares are compatible with the simulation for $\tau = 5$, but longer ones agree better with $\tau = 10$ and $\tau = 15$. We conclude that, sticking to the single $\tau$ simplification, $\tau = 10$ hours gives the best agreement with both the temporal length of flares and with their frequencies. Nevertheless, we are aware that the assumption of a single value for $\tau$ for all flares is only a simplification and does not reflect the real behavior of the sources. The possible discrepancy in the flare frequency of the brightest sources may indicate that for these stars, dominated by intense flares, $\tau$ is on average longer. Of course, similar effects may be reproduced by more complicated models: e.g. a dependence of $\alpha$ on flare energy would probably produce a similar effect.

## 5. Discussion

### 5.1. Low Mass vs. Solar Mass Stars

In Sect. 3.1, we compared the frequencies of flares for low mass and for solar mass stars, finding that, at ∼ 1 Myr there is no appreciable difference. Variability studies, conducted with a non parametric method on ∼ 100 Myr old Pleiades members (Marino et al., 2000, 2003) have shown that old low mass stars are more X-ray variable then higher mass ones. Our results therefore indicate that variability evolves differently in the two mass regimes: while low mass stars keep a high level of variability, solar mass stars become relatively less variable as they age. We note that a similar trend is also observed for the average X-ray luminosity, when measured relatively to the bolometric stellar luminosity (e.g. Flaccomio et al., 2003).

Our method to detect flares and compare the variability for stars in different ranges of mass also permits the analysis of weak sources, that are impossible to study with non parametric methods. Our analysis indeed can be applied to all the stars and, by means of a relation between the expected number of flares and the intensity of the source, it is able to account for the dependence of the flare detection sensitivity on source statistics.

The fundamental difference in the evolution of solar mass and low mass stars between 1 and 100 Myr is that solar mass stars change radically in internal structure while low mass stars do not. At 1 Myr stars have a fully convective structure in both mass ranges; as G type stars evolve, a radiative core develops, while low mass stars remain fully convective. This different evolution may be responsible for the observed differences in the evolution of X-ray activity and in particular on the flare variability. More specifically, since the X-ray activity is related to the magnetic field of the star and to its angular momentum (Pallavicini et al., 1981), the different variability evolution may be related to the different rotational evolution of solar and lower mass stars in the PMS and early MS phases. Angular momentum losses are in fact higher in the surface layers of solar mass stars because of the decoupling between the convective envelope and the radiative core (Bouvier, 1997; Bouvier et al., 1997). At 1 Myr, however, the internal structure is still the same in both mass ranges, and this could be reflected by similar X-ray activity and variability.

Other explanations of the different evolution of the variability properties of G and M type stars may involve the nature of the dynamo. Barnes (2003) interpreted the measured rotational periods of solar and late type stars through a model in which young fully convective stars are dominated by a turbulent dynamo; as solar mass stars evolve, a shear between the radiative and a convective zones develops, leading to the development of an interface dynamo; M type stars that remain fully convective do not possess such an interface so that the X-ray activity would remain related to the turbulent dynamo.

However, although M type stars remain quite variable between 1 and 100 Myr, there seems to be is an evolution of the timescales of flares. We found that the mean total duration of flares in the low mass sample is 21 hours (compatible with an e-folding time of 10 hours); studies of older M stars (see e.g. study of the AD Leo by Güdel et al., 2003) found significantly shorter flares with typical duration of 3.6 ks in the EUV and 0.3-0.4 ks in X-rays. The time scale of flares is related to the size of the confining magnetic structures and to the presence of sustained heating (see Reale 2002 for an extensive review): according to the scaling relation law in Serio et al. (1991) the typical size of the flaring loops in our sample is ∼ 100 times larger than in older M stars, if no sustained heating is present. Young low mass stars likely have considerably larger active regions than more evolved stars and flaring loops with sustained heating during the flare decay. Indeed, from a detailed analysis of 32 of the most powerful flares in the COUP dataset, Favata et al. (2005) found that some flaring structures may be even larger than the radius of the stars.

### 5.2. Continuous Flaring and Microflaring

The stars of the low mass sample in the ONC appear quite variable, with several flares of different intensity, superimposed on an approximatively constant emission level. We have shown that the light curves are consistent with a model in which the emission is completely due to flares, with a power law energy distribution. We are able to detect only the intense events, while the weak flares merge and form a *pseudo-quiescent* level, that we call the characteristic level.

In our simple model the power law has the same slope for all the stars, while the characteristics of each light curve are determined by a combination of the total number of flares emitted during the observation ($N_{fl}$) and the minimum allowed number of counts in a flare ($C_{min}$). This latter parameter accounts for most of the variability, and, since we observe a large range of different behaviours in the light curves (see Fig. 2), its values cover a large range. Note that the different values of $C_{min}$ for the flare intensity of individual sources, also contribute to the turnover of the CDF of the sample at low counts. However, we cannot positively attribute a physical meaning to $C_{min}$. Our analysis does not indeed exclude more complicated models, given that the intensity distribution of undetected flares is effectively unconstrained by the data. For example, a constant quiescent emission, although unnecessary in our model, could still be present and would be indistinguistanble from the time integrated effect of small flares. Our $C_{min}$ values would therefore represent just lower limits to the intensity of the smallest flares.

Theories that try to explain the coronal heating only with flares and microflares have been proposed originally for the solar corona (Lin et al., 1984; Porter et al., 1987; Parker, 1988). Although the energy released by the single microflare is on average small, the energy distribution could be steep enough to make the total energy released by small flares comparable to the energy needed for the heating of active regions and even of the quiet corona. Although solar studies found $\alpha$ values on the order of $1.6 - 1.8$ for solar microflares (Hudson, 1991), more recent studies with higher spatial and spectroscopic resolution have found that different slopes apply in different energy intervals (e.g. Shimizu & Tsuneta, 1997; Parnell & Jupp, 2000). The power law may steepen at lower energies to $\alpha > 2$ and flares could suffice to heat all the solar corona. In our case, the situation is quite different because the energies of the events under study are much larger. Making the reasonable assumption that the energy of detected photons is ∼ 1 keV, we estimate that the smallest flares required by our model (0.001 counts) have energies of the order of $10^{29}$ erg, i.e. two order of magnitudes more energetic than solar microflares; indeed, in our sources, a solar-like corona, if it exists at all, would remain unobserved; nevertheless, it is interesting that in both cases we find that the coronal emission could be explained by a continuous distribution of flares.

Other studies have also suggested that the whole X-ray emission of active stars could be due to coronal flares. Guedel (1997) demonstrated that the double peaked time-averaged emission measure observed in active stars can be explained by flare emission, suggesting that the flare-heating model is a good candidate for the identification of the main coronal heating mechanism. Güdel et al. (2003) studied the flare statistics of a long observation of AD Leo, using both spectral and temporal information, and found $\alpha$ in the interval $2.0 - 2.5$. Moreover, Telleschi et al. (2005) measured $\alpha$ from the slope of the high temperature tail of the differential emission measure of six nearby main sequence G stars; they found $\alpha = 2.2 - 2.8$, compatible with the hypothesis that most of the coronal heating is due to flaring.

## 6. Summary and Conclusion

We studied the short term variability of a low mass sample ($0.1 - 0.3 M_\odot$) of COUP sources, with the purpose of characterizing the flare properties of these stars. We quantified the variability of 165 light curves by means of Maximum Likelihood Blocks and established an operative definition of flares in order to detect them. Our method permits an unbiased study of large stellar samples and we plan to use it for other subsamples of COUP sources (e.g. different mass ranges) and for stars in other regions. However the method is more effective when applied to long and sensitive observations of large stellar samples: a long exposure time is needed in order to establish the characteristic level for each light curve and, together with the sample size, to observe a number of statistically significant flares; high sensitivity is also needed to detect faint events so to better determine the low energy part of the flare energy distribution.

Our main results can be summarized as follows:

– We have determined the flare frequency as function of source statistics. We find that the flare frequencies in solar and low mass stars are indistinguishable.
– The high energy tail of the distribution of flare energies is compatible with a power law with a slope $\alpha \sim 2.2$. We have tested through extensive simulation the hypothesis that the light curves are formed entirely by flares having power law energy distributions with an universal slope. We have determined for each source the free parameters of the model, i.e. the total number of flares and the low energy cutoff of the distributions. Our simple model successfully reproduces several observed characteristics of our light curves: the mean and characteristic emission levels, the number of detected flares, the mean length of flares and the distribution of detected flare intensities.

*Acknowledgements.* We acknowledge financial contribution from contract ASI-INAF I/023/05/0 and from PRIN-INAF 2005. E.D. Feigelson is funded by Chandra grants SAO GO3-4009A and NAS8-38252.

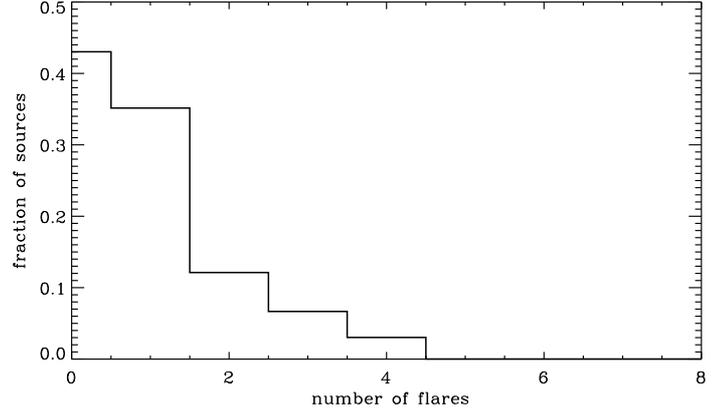

**Fig. 3.** Distribution of the number of flares per source in our sample. Note that 40% of our stars have no flares during the observation, but there are sources in the same sample with several flares.

**Table 1.** Frequency of flares for subsamples of stars with different minimum number of counts

| minimum counts | sources in the sample | frequency of flares per source [ks$^{-1}$] | mean flares per source* |
|---|---|---|---|
| 100 | 165 | 1/916 | 0.9 ± 0.08 |
| 200 | 141 | 1/809 | 1.03 ± 0.09 |
| 500 | 99 | 1/664 | 1.26 ± 0.12 |
| 1000 | 70 | 1/592 | 1.41 ± 0.14 |
| 2000 | 43 | 1/538 | 1.56 ± 0.18 |
| 3000 | 26 | 1/444 | 1.88 ± 0.24 |
| 4000 | 16 | 1/406 | 2.06 ± 0.29 |

\* the error is the standard deviation of the mean

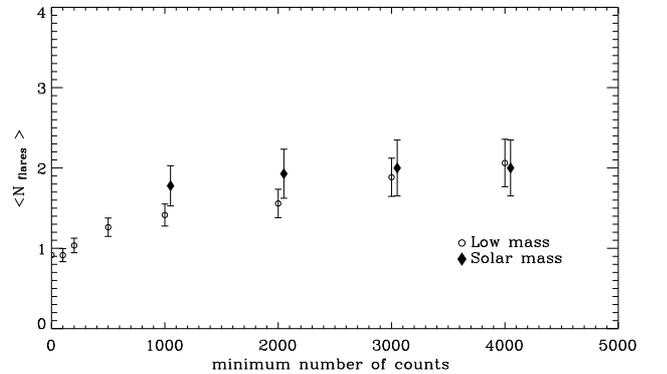

**Fig. 4.** Mean number of flares per sources for subsamples with total source counts larger than a given value. Circles refer to the low mass stars, diamonds to the solar mass stars (Wolk et al., 2005). In this plot solar mass data are slightly shifted to the right for sake of clarity.

**Table 2.** Model parameters used to generate a grid of models for the exploratory simulations.

| parameters | values |
|---|---|
| $\alpha$ | 2.1, 2.2, 2.4 |
| $C_{min}$ | 0.1, 1, 10 |
| $<N_{fl}>$ | 500, 1000, 2000 |
| $\tau$ (hours) | 5, 10 |

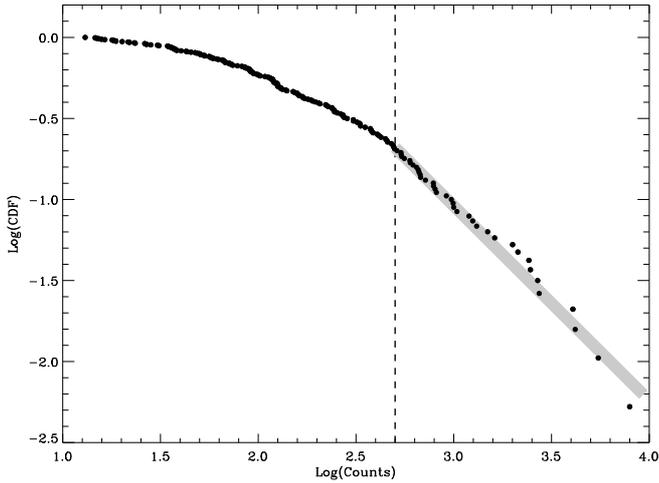

**Fig. 5.** Cumulative distribution function of flare intensities (CDF); for high counts, to the right of the dashed line, the distribution is compatible with a power law with $\alpha - 1 = 1.2 \pm 0.2$, shown as a solid grey line; for low counts the distribution appears to saturate, most likely because the detection of low-counts flares is incomplete.

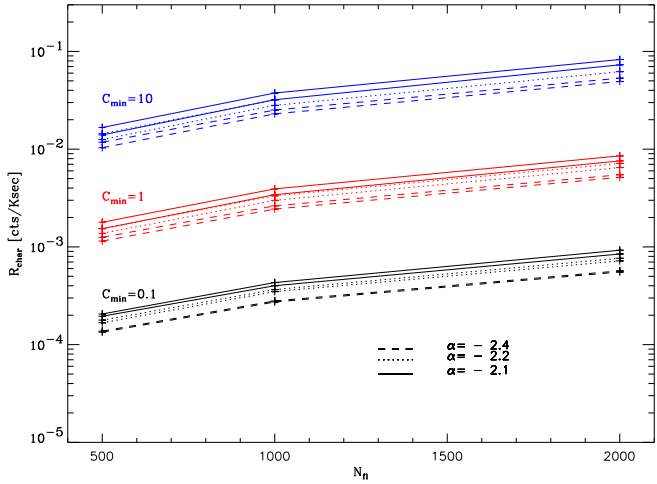

**Fig. 6.** Dependence of $R_{char}$ on the model parameters. $R_{char}$ increases as we increase $C_{min}$ and $\alpha$. The couples of curves with the same type of line are relative to different values of $\tau$.

**Table 3.** Mean temporal length of flares for simulations with different $\tau$.

| $\tau$ | mean temporal length of flares (hours) |
|---|---|
| 5 | 15 |
| 8 | 18 |
| 10 | 21 |
| 12 | 23 |
| 15 | 27 |

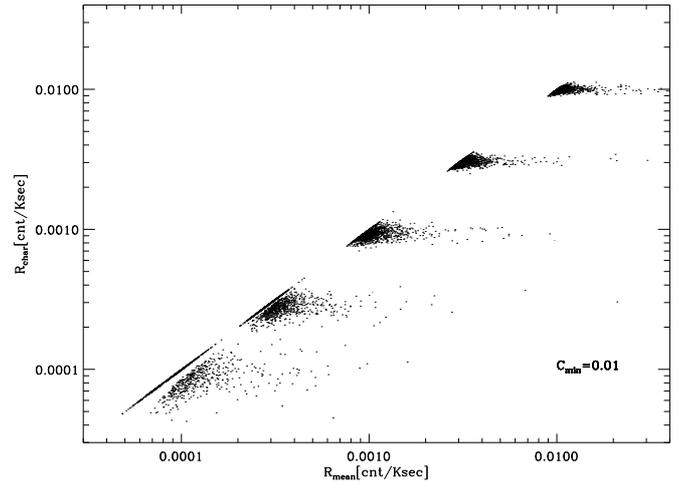

**Fig. 7.** $R_{char}$ vs $R_{mean}$ for a set of five simulations with the same $C_{min}$ and different $< N_{fl} >$. Each point represents a result of a single simulation. The five samples, referring to five values of $< N_{fl} >$, are easy distinguishable because $R_{mean}$ and $R_{char}$ increase with $< N_{fl} >$. The gaps inside the clouds of points for low $R_{mean}$ are due to the discretization in the MLB algorithm and to the thresholds used in our definitions of the characteristic level.

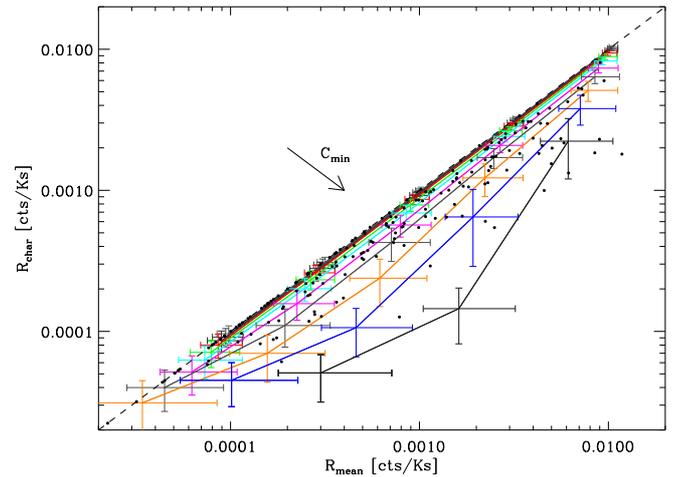

**Fig. 8.** Synthesis of the simulation results. The crosses represent the median $R_{mean}$ and the mean $R_{char}$ for each simulation set, with their 68% quantiles. The connected crosses refer to the same value of $C_{min}$. The dots represent real sources.

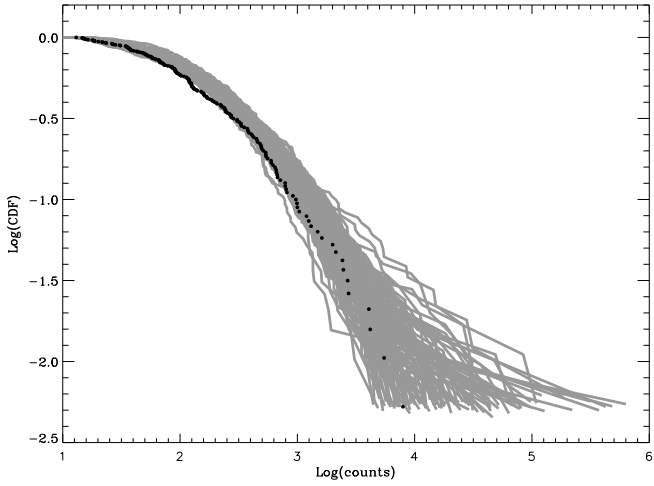

**Fig. 10.** CDF distributions for the real sources (black dots) and for the 100 simulations of the same sample (grey lines). Although our model is very simple (e.g. we use just one $\tau$), our sources appear to be compatible with one of the possible realizations.

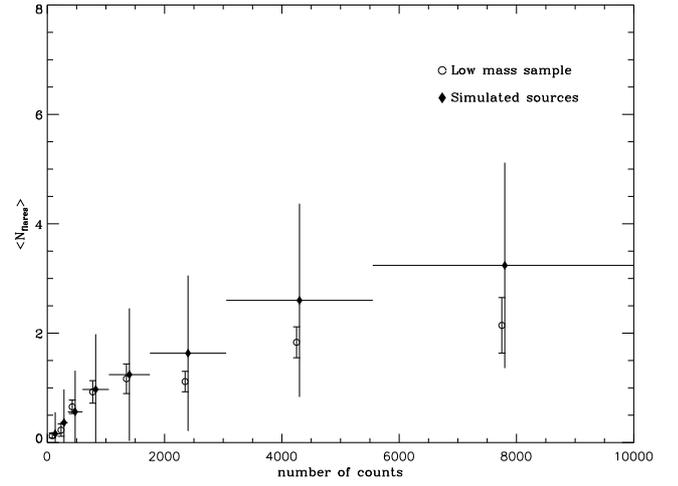

**Fig. 12.** Mean number of flares per source, detected in the low mass sample and in the simulations (see legend). Horizontal bars represent the count interval in which the mean values are calculated. Vertical bars are the 1 $\sigma$ dispersions for the simulation results and the standard error of the mean for the real stars.

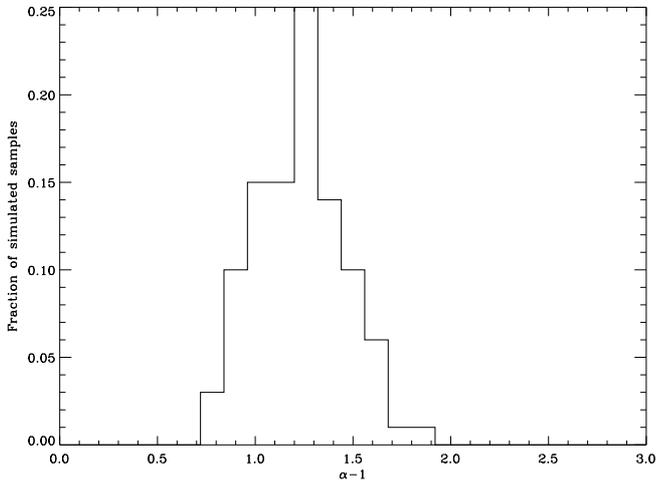

**Fig. 11.** Distribution of the power law slopes of the CDFs obtained from the 100 synthetic samples. The distribution has a mean value of $\alpha - 1 = 1.2 \pm 0.2$, equal to the input value of the simulations, calculated from the data.

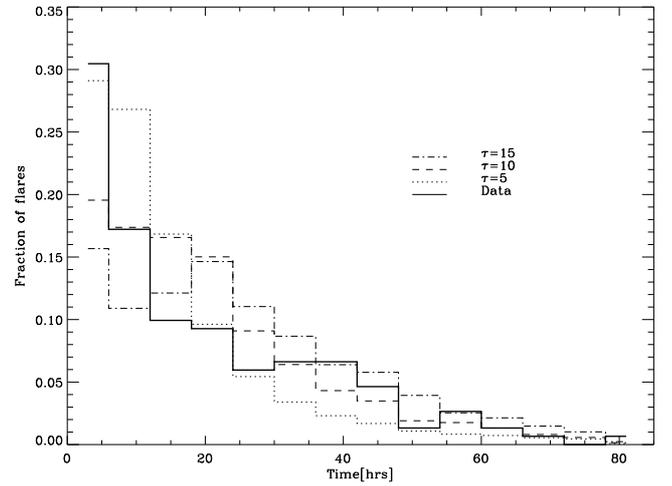

**Fig. 13.** Histograms of the duration of flares for the data and three simulations with different input values of $\tau$. Different types of line are used for the four distributions, as shown in the legend.